\documentclass[aps,prl,twocolumn,fleqn]{revtex4-1}

\usepackage{amsmath, amssymb}
\usepackage[pdftex]{graphicx}
\usepackage{color}

\begin{document}

\title{High Baryon Densities in Heavy Ion Collisions at Energies Attainable at the BNL Relativistic Heavy Ion Collider and the CERN Large Hadron Collider}

\author{Ming Li and Joseph I. Kapusta}
\affiliation{School of Physics and Astronomy, University of Minnesota,
Minneapolis, Minnesota 55455, USA}

\begin{abstract}
In very high energy collisions nuclei are practically transparent to each other but produce very hot, nearly baryon-free, matter in the so-called central rapidity region.  The energy in the central rapidity region comes from the kinetic energy of the colliding nuclei.  We calculate the energy and rapidity loss of the nuclei using the color glass condensate model.  This model also predicts the excitation energy of the nuclear fragments.  Using a space-time picture of the collision we calculate the baryon and energy densities of the receding baryonic fireballs.  For central collisions of gold nuclei at the highest energy attainable at the Relativistic Heavy Ion Collider, for example, we find baryon densities more than ten times that of atomic nuclei over a large volume.    
\end{abstract}

\maketitle

In 1980 Anishetty, Koehler and McLerran \cite{AKM1980} outlined what happens when large nuclei collide at extremely relativistic energies.  Rather than stopping, as at lower energies, the nuclei pass through each other, compressing and depositing energy in each other.  Most of the produced particles appear in the region between the two receding nuclei, the so-called inside-outside cascade.  They argued that the matter within these fireballs would quickly thermalize with a baryon density about 3.5 times that of atomic nuclei, and an energy density of about 2 GeV/fm$^3$, which is four times the energy density of a proton.  In a seminal paper in 1983 Bjorken \cite{BJ1983} utilized the inside-outside cascade to propose a hydrodynamic model for the evolution of the matter produced in the central rapidity region between the receding fireballs.  Ever since then the community has been focused on the central region because (i) the energy density is expected to be higher there, (ii) the matter is nearly baryon-free, making it more relevant for the type of matter that existed in the early universe, and (iii) detectors in a collider can more easily measure particle production and correlations in a few units of rapidity around the center of momentum.  In the ensuing decades only a few papers have studied the baryonic fireballs and their deceleration \cite{C1984,GC1986,MK2002,FS2003,ML2007}.  In this paper we use the McLerran-Venugopalan model \cite{MV1994} to compute the rapidity loss and excitation energy of the fireballs followed by a space-time picture to obtain the energy and baryon densities.  Currently there is much interest in studying high baryon density matter in the laboratory \cite{CBM}, albeit at center of momentum energies much lower than discussed in this paper.

Consider central collisions of equal mass nuclei; this is easily relaxed to noncentral collisions and collisions of unequal mass nuclei.  We neglect transverse motion, which should not be important during the fraction of a fm/c time interval of relevance.  Then the collision can be thought of as a sum of independent slab-slab collisions each taking place at a particular value of the transverse coordinate ${\bf r}_\perp$ with the beam along the $z$-axis.  The projectile slab has a four-momentum per unit area in the center-of-momentum frame denoted by ${\cal P}_{\rm P}^{\mu} = ({\cal E}_{\rm P}, 0, 0, {\cal P}_{\rm P})$.  The slab loses energy and momentum to the classical color electric and magnetic fields produced in the region between the two receding slabs, sometimes called glasma.  This loss is quantified by $d{\cal P}^{\mu}_{\rm P} = -T_{\rm glasma}^{\mu\nu}d\Sigma_{\nu}$ where $d\Sigma_{\nu} = (dz,0,0,-dt)$ is the infinitesimal four-vector perpendicular to the hypersurface spanned by $dt$, $dz$, and unit transverse area.  The energy-momentum of the glasma has the form \cite{CFKL2015,LK2016}
\begin{equation}\label{em_tensor}
T^{\mu\nu}_{\rm glasma}=
\begin{pmatrix}
\mathcal{A}+\mathcal{B}\cosh{2\eta} & 0 & 0 & \mathcal{B}\sinh{2\eta} \\
0 & \mathcal{A} & 0 & 0 \\
0 & 0 & \mathcal{A} & 0 \\
\mathcal{B}\sinh{2\eta} & 0 & 0 & -\mathcal{A}+\mathcal{B}\cosh{2\eta} \\
\end{pmatrix} \, .
\end{equation}
The $\mathcal{A}$ and $\mathcal{B}$ are known analytical \cite{LK2016} and numerical (for SU(2)) \cite{Gelis:2013rba} functions of proper time $\tau =\sqrt{t^2-z^2}$ (and other input parameters), while the dependence on space-time rapidity $\eta=\frac{1}{2}\ln [(t+z)/(t-z)]$ follows from the fact that $T^{\mu\nu}_{\rm glasma}$ is a second-rank tensor in a boost-invariant setting.  The longitudinal position of the slab $z_{\rm P}$ is a function of time, $z_{\rm P}=z_{\rm P}(t)$.  The $z_{\rm P}$ is related to the time $t$ via the velocity $v_{\rm P} = dz_{\rm P}/dt =\tanh{y_{\rm P}}$, where $y_{\rm P}$ is the momentum-space rapidity of the slab. So all the quantities solely depend on $t$.  Of course $T^{\mu\nu}_{\rm glasma}$ must be evaluated on the trajectory of the slab.  Explicitly
\begin{equation}\label{eom}
\begin{split}
&d{\cal E}_{\rm P}(t,z_{\rm P}) = -T^{00}_{\rm glasma}(t,z_{\rm P})dz_{\rm P} + T^{03}_{\rm glasma}(t,z_{\rm P})dt \\
&d{\cal P}_{\rm P}(t,z_{\rm P}) = -T^{30}_{\rm glasma}(t,z_{\rm P})dz_{\rm P} + T^{33}_{\rm glasma}(t,z_{\rm P})dt \,. \\
\end{split}
\end{equation}
It is useful to define the Lorentz invariant effective mass per unit area ${\cal M}_{\rm P}$ via the relations ${\cal E}_{\rm P} = {\cal M}_{\rm P} \cosh{y_{\rm P}}$ and ${\cal P}_{\rm P} = {\cal M}_{\rm P} \sinh{y_{\rm P}}$.  The above pair of equations (\ref{eom}) describe not only the loss of kinetic energy of the projectile nucleus but also the internal excitation energy imparted to it during the collision.  Thus ${\cal M}_{\rm P}$ is not constant but increases with time, unlike the case of the string model \cite{MK2002}; this difference can be traced to the lack of off-diagonal terms in the energy-momentum tensor representing the strings but which are present in (\ref{em_tensor}). The thickness of the glasma slice at $\tau=0$ is zero and, since the energy density in the glasma is finite at $\tau=0$, the total energy initially in the glasma is zero.

Initial conditions are needed to solve the equations of motion.  Immediately after the nuclei collide at $\tau = 0$ the local energy density in the glasma is \cite{FKL2006,Lappi:06,Fujii:2008km,CFKL2015}
\begin{equation}
\varepsilon_0 (r_{\perp}) = \frac{2\pi  N_c \alpha_s^3}{N_c^2-1} \mu^2 (r_{\perp},Q) \ln^2(Q^2/\Lambda_{\rm QCD}^2) \, . 
\label{initial}
\end{equation}
Here $\alpha_s$ is the (fixed) strong coupling, $\mu$ is the width of a Gaussian which characterizes color charge fluctuations at transverse distance $r_{\perp}$, and $Q$ is an ultraviolet cutoff on transverse momentum which characterizes the division between the classical gluon fields and perturbative QCD.  Larger values of $Q$ attribute more energy and momentum to the classical fields while smaller values of $Q$ attribute more to production of partons or minijets.  The value of $Q$ should be chosen optimally so that obervable results are minimally sensitive to it.  In what follows we choose $3 \le Q \le 5$ GeV with a favored value of 4 GeV.  Generally $\mu(r_{\perp})$ is taken to be proportional to the thickness function $T_{\rm A}(r_{\perp}) = \int_{-\infty}^{\infty} dz \rho_{\rm A}(r_{\perp},z)$ where $\rho_{\rm A}$ is the nucleon number density of a nucleus of atomic number A.  We follow this practice using a Woods-Saxon distribution for the nucleus.

Equation (\ref{initial})  contains significant uncertainties via the numerical values of $\alpha_s$ and $\mu (r_{\perp},Q)$.  For the absolute normalization, therefore, we turn to hydrodynamical descriptions of collisions at the top RHIC energy of $\sqrt{s_{NN}} = 200$ GeV.  Reference \cite{SBH2011} assumed that viscous hydrodynamics became applicable at $\tau = 0.6$ fm/c with $\varepsilon (r_{\perp}=0, \tau = 0.6 \; {\rm fm/c}) = 30$ GeV/fm$^3$.  Extrapolating back to $\tau = 0$ using the results in Ref. \cite{LK2016} gives $\varepsilon_0 \equiv \varepsilon_0 (r_{\perp}=0,\tau=0) =$ 123, 142, and 158 GeV/fm$^3$ for $Q=$ 3, 4 and 5 GeV, respectively.  Other analyses result in somewhat higher values of the energy density at $\tau = 0.6$ fm/c \cite{SHHS2011}; these would increase the rapidity and energy loss, making our conclusion even stronger.

\begin{figure}[tb]
\includegraphics[width=0.99\columnwidth]{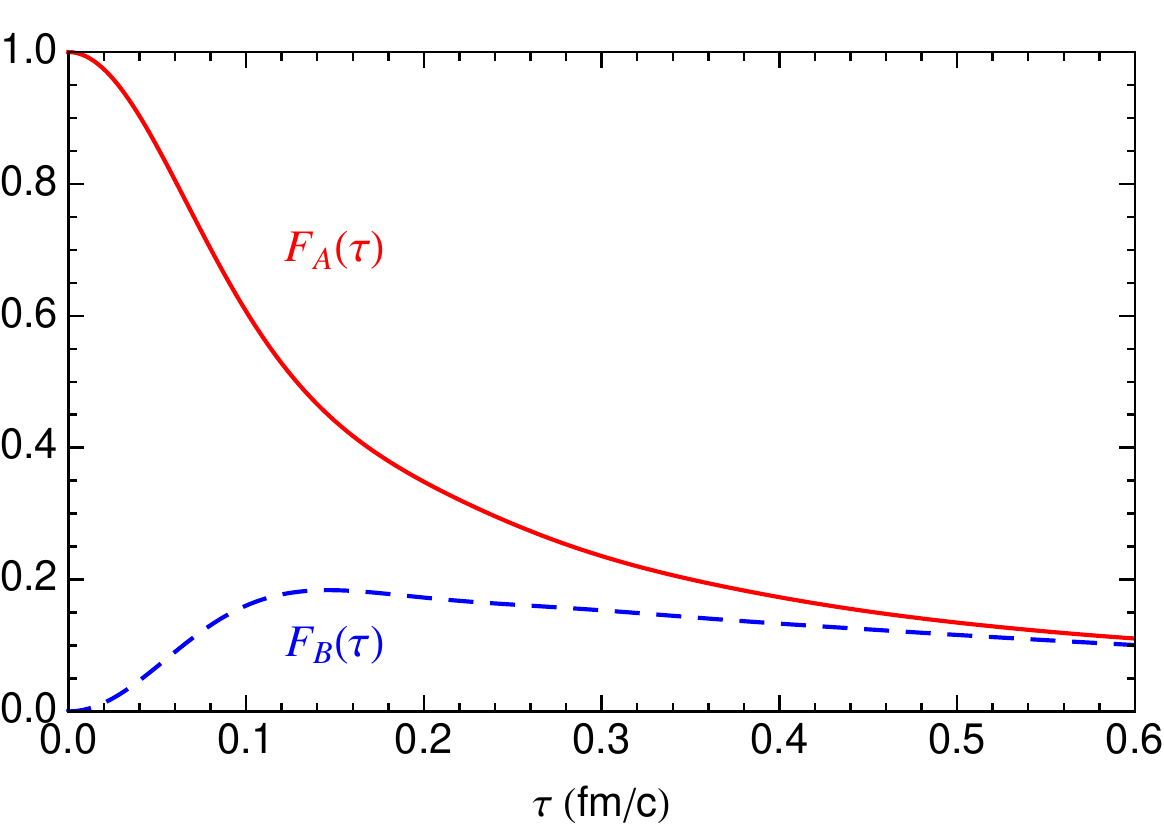}
\caption{(color online) The dependence of $F_{\cal A}$ and $F_{\cal B}$ on proper time for $Q = 4$ GeV.}
\label{fig:1}
\end{figure}

The function
\begin{equation}
{\cal A} = \varepsilon_0 \, [T_{\rm A}(r_{\perp})/T_{\rm A}(0)]^2 \, F_{\cal A}\left(\ln(Q^2/\Lambda_{\rm QCD}^2), Q\tau\right)
\end{equation}
is now fixed.  The dimensionless function $F_{\cal A}$ can be found in Ref. \cite{LK2016}.  The function ${\cal B}$ has the same form but with a different $F_{\cal B}$.  The functions 
$F_{\cal A}$ and $F_{\cal B}$ are plotted in Fig. \ref{fig:1} for $Q = 4$ GeV. 

Now it is just a matter of solving the pair of equations \eqref{eom} for ${\cal M}_{\rm P}$ and $y_{\rm P}$  numerically.  We solve them up to $\tau = 0.6$ fm/c where it is assumed that a transition from glasma to quark-gluon plasma has taken place \cite{SBH2011}.

\begin{figure}[bt]
\includegraphics[width=0.99\columnwidth]{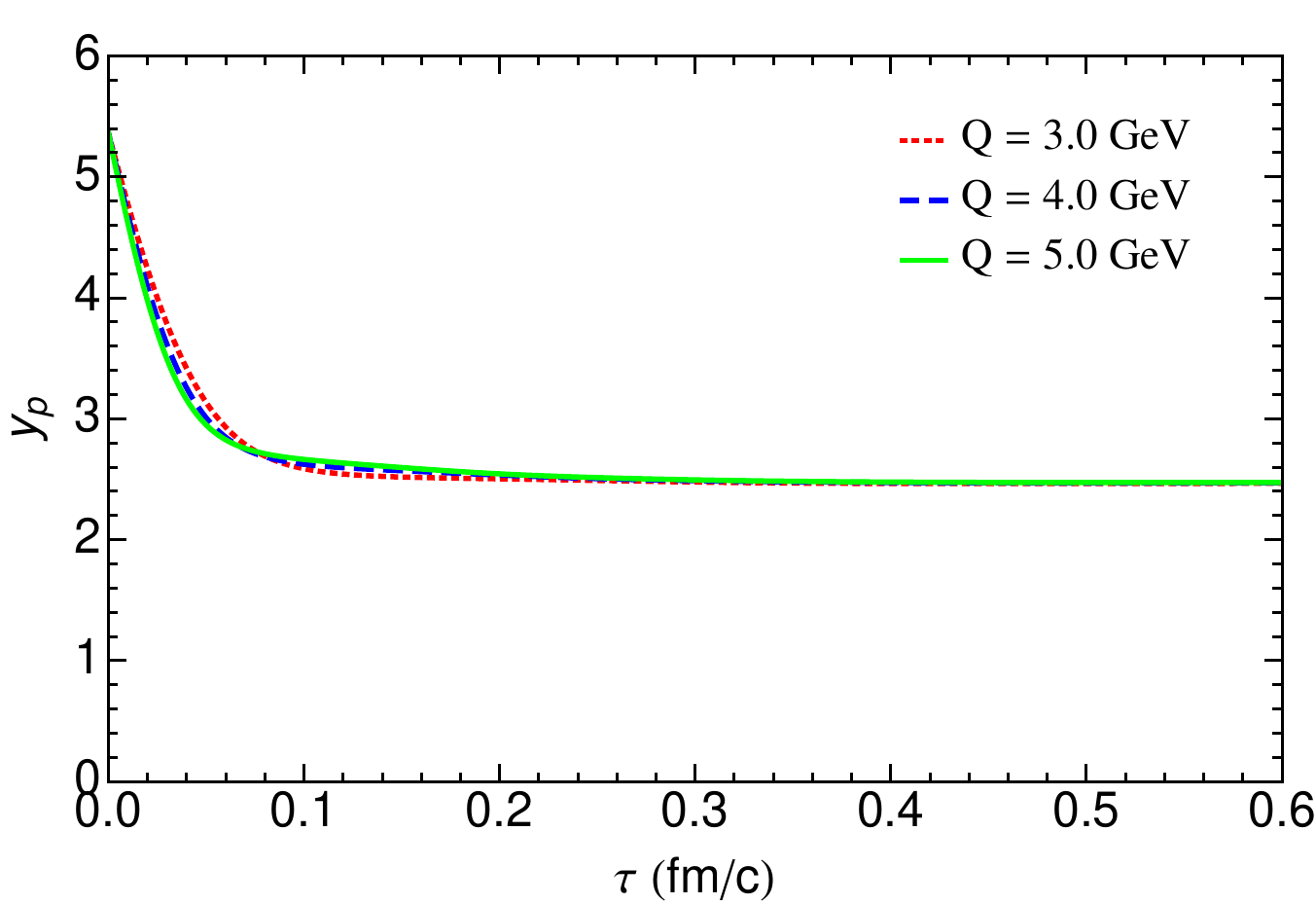}
\caption{(color online) Rapidity of the central core of a Au projectile nucleus in the center-of-momentum frame for $\sqrt{s_{NN}} = 200$ GeV as a function of proper time.  The result is insensitive to the choice of $Q$ in the physically relevant range.}
\label{fig:2}
\end{figure}
Figure \ref{fig:2} shows the momentum-space rapidity $y_{\rm P}$ of the central core of a gold nucleus as a function of proper time $\tau$.  (The beam rapidities in the center-of-momentum frame are $\pm 5.36$.)  The central core loses about three units of rapidity within the first 0.1 to 0.2 fm/c; this is a robust result, insensitive to the value of $Q$.  When averaged over the whole nucleus the baryon rapidity loss is about 2.4.  BRAHMS \cite{BRAHMS2004,BRAHMS2009} was the only detector at RHIC or LHC that could measure particle production anywhere near the fragmentaion region.  The coverage was limited to $y \le 3.1$, so the uncertainty in the loss estimate was large.   For 0-10\% centrality BRAHMS found an average rapidity loss of about $2.05 +0.4/-0.6$.  This is consistent with our result, especially since we focus on 0\% centrality for illustration.

\begin{figure}[ht]
\includegraphics[width=0.99\columnwidth]{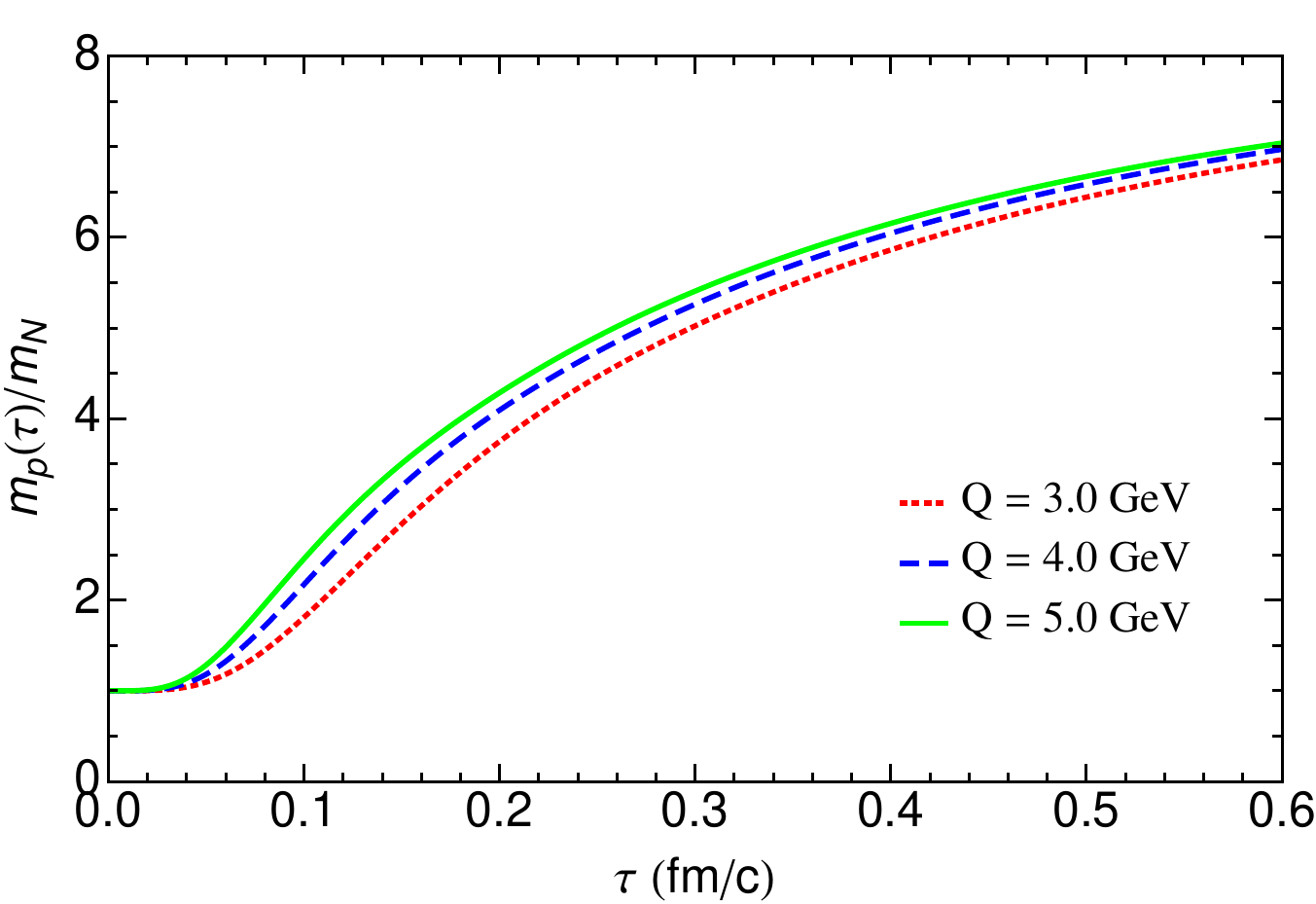}
\caption{(color online) Excitation energy per baryon in the central core of a Au projectile nucleus in the center-of-momentum frame for $\sqrt{s_{NN}} = 200$ GeV as a function of proper time.  The result is mildly sensitive to the choice of $Q$ in the physically relevant range.}
\label{fig:3}
\end{figure}

Figure \ref{fig:3} shows the excitation energy per baryon in units of the nucleon mass as a function of proper time.  There is a slow but monotonic increase, unlike the rapidity loss whose asymptotic limit is reached within a few tenths of a fm/c.  There is a weak dependence on $Q$.

The McLerran-Venugopalan model assumes that the projectile and target nuclei move along the light-cone.  The validity of the model increases with increasing beam energy.  It assumes that the nuclei can be treated as infinitesimally thin slabs.  This is valid to a high degree of accuracy, but it does not address the space-time evolution of the individual nuclei.  (A modification of the model to give the nuclei a nonzero thickness was performed in Ref. \cite{Lam} and its parameters were determined in Ref. \cite{Ozonder}.  However, the space-time evolution of the glasma was not found and so cannot be used here.)  To estimate the space-time evolution requires additional input.  Anishetty {\it et al.} \cite{AKM1980} presented a very simple and direct calculation that the nuclear matter would be compressed in its own rest frame by a factor of $\exp(\Delta y)$ where $\Delta y > 0$ is the rapidity loss (gain) of the projectile (target).  It is clearly a Lorentz invariant quantity which follows from the infinitely thin projectile sweeping through the target in the target rest frame.  The argument was verified in a specific model in Ref. \cite{GC1986}.  Noting that $\Delta y$ depends on the transverse coordinate $r_{\perp}$ the local proper baryon density is
\begin{equation}
n_B(r_{\perp},z') = {\rm e}^{\Delta y(r_{\perp})} \rho_{\rm A}\left(r_{\perp}, z'  {\rm e}^{\Delta y(r_{\perp})}\right)
\end{equation}
where $z' = z - z_{\rm P}(r_{\perp})$, all evaluated at $\tau = 0.6$ fm/c. 
\begin{figure}[t]
\includegraphics[width=0.999\columnwidth]{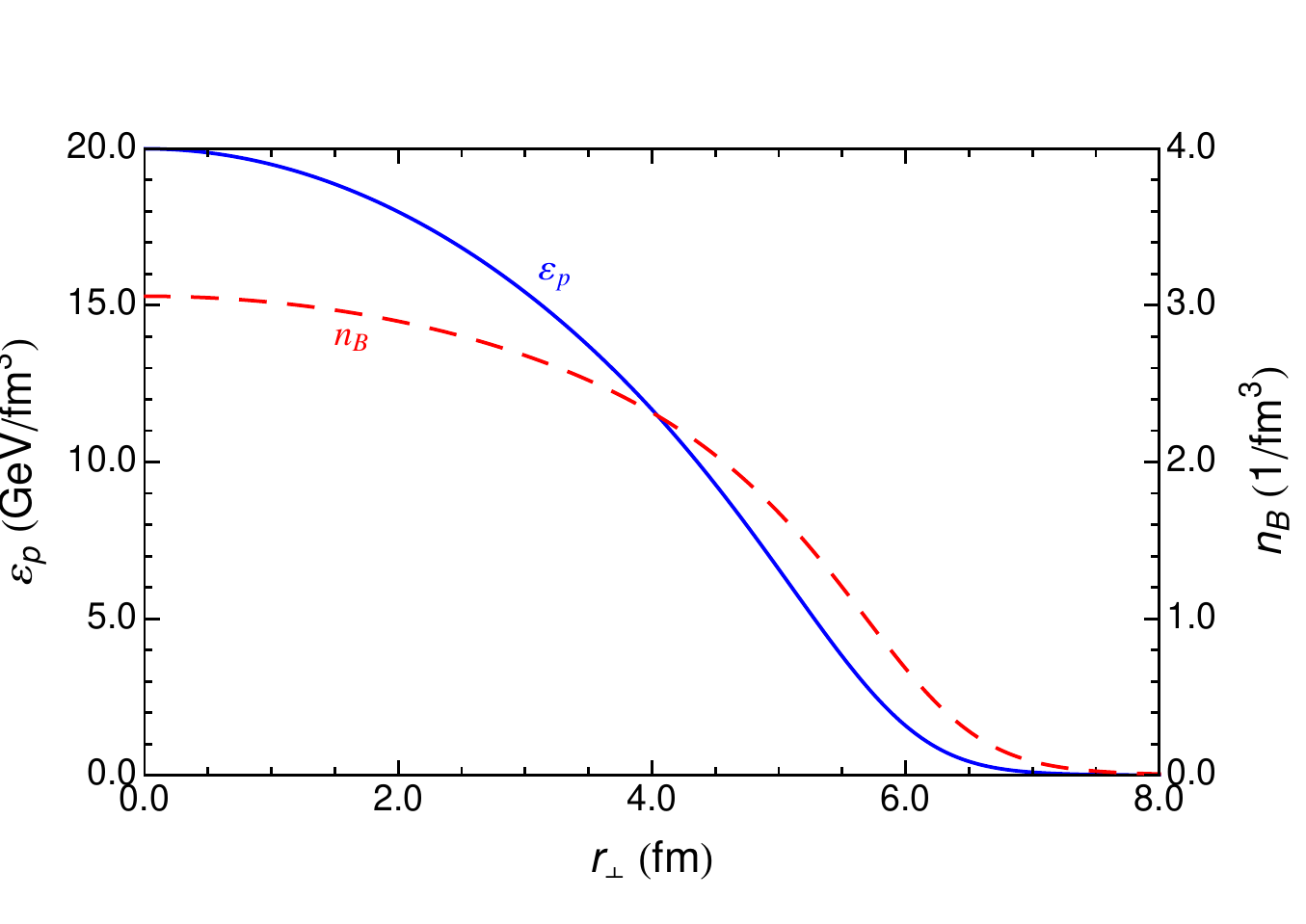}
\caption{(color online) The energy density and baryon density at $\tau = 0.6$ fm/c as functions of the transverse distance for central collisions of Au nuclei at $\sqrt{s_{NN}} = 200$ GeV.}
\label{fig:4}
\end{figure}

Figure \ref{fig:4} shows the proper energy density and baryon density as functions of the transverse coordinate for $Q = 4$ GeV at $\tau = 0.6$ fm/c.  As can be seen from the previous figures, the baryon density is less sensitive to the time at which the transition from glasma to quark-gluon plasma occurs than the excitation energy.  It should be noted that the maximum baryon density, about 3 baryons/fm$^3$, is 20 times greater than the normal matter density of 0.155 nucleons/fm$^3$. 

\begin{figure}[hb!]
\includegraphics[width=0.99\columnwidth]{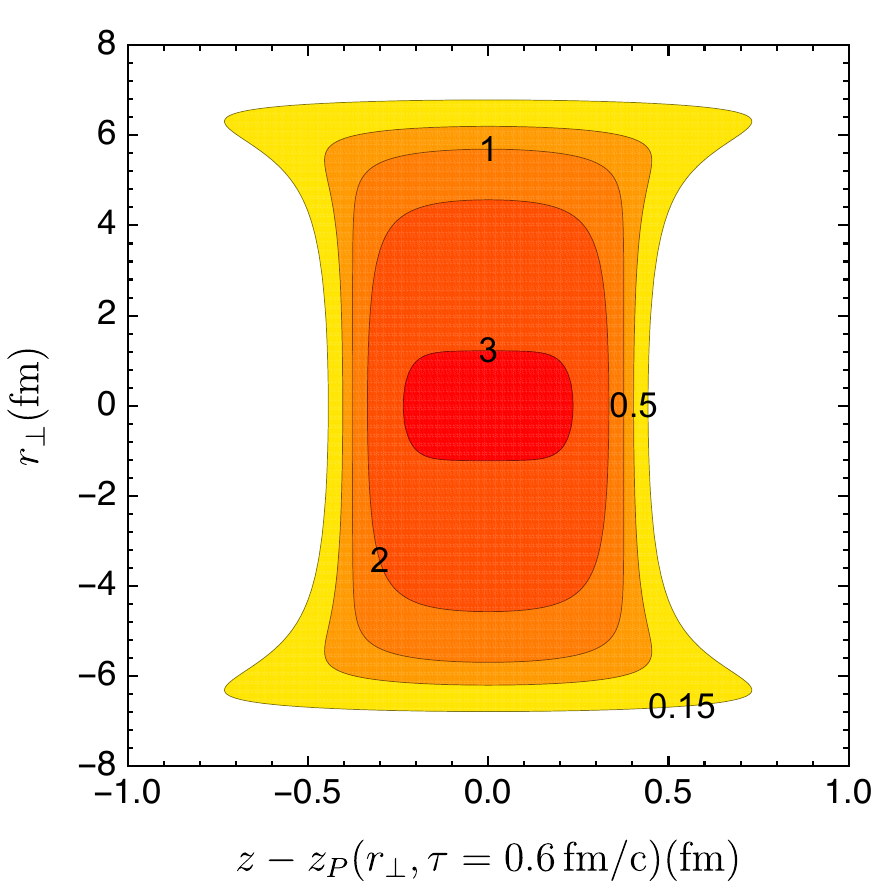}
\caption{(color online) Contour plot of the proper baryon density for central collisions of Au nuclei at $\sqrt{s_{NN}} = 200$ GeV.  The numbers are in units of baryons per fm$^3$.  The horizontal axis measures the distance along the beam direction in the local rest frame.  Care must be taken when interpreting this plot since the rapidity of the matter, and therefore the frame of reference, depends on $r_\perp$}.
\label{fig:5}
\end{figure}
Figure \ref{fig:5} is a contour plot of the proper baryon density.  The contours are drawn at $n_{\rm B} = 3, 2, 1, 0.5$, and 0.15 baryons/fm$^3$.  The shapes of the contours arise for the following reasons.  The diameter of a gold nucleus $2R_{\rm A}$ is about 14 fm.  The core centered at $r_{\perp} = 0$ along the $z$-axis contains the most matter, suffers the greatest deceleration, and hence the greatest compression.  Moving outward with increasing $r_{\perp}$, the length of the tube is decreased to $2\sqrt{R_{\rm A}^2 - r_{\perp}^2}$ and the deceleration, and hence compression, are reduced.  These opposing effects approximately cancel each other, giving rise to roughly rectangular contours in the $r_{\perp}-z$ plane.  Care must be taken when interpreting this figure.  Since the rapidity loss depends on $r_{\perp}$ it means that there is a shear in the $r_{\perp}$-direction and there is no single, global frame of reference for all elements of the fireball.  

\begin{figure}[hb]
\includegraphics[width=0.99\columnwidth]{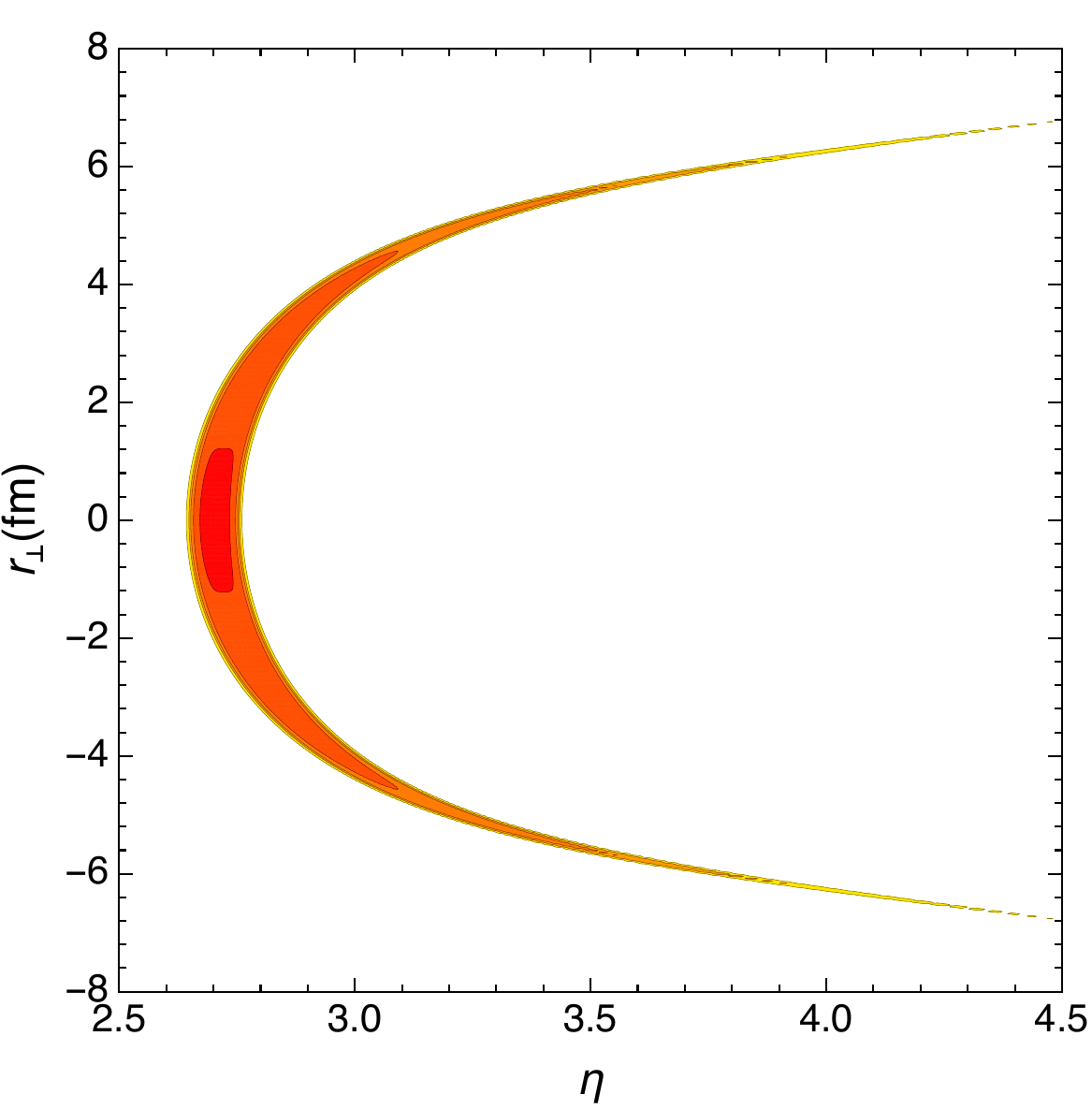}
\caption{\label{F4}(color online)  Contour plot of the proper baryon density at $\tau = 0.6$ fm/c for central collisions of Au nuclei at $\sqrt{s_{NN}} = 200$ GeV.  The units are the same as in Fig. 5.  The horizontal axis is the space-time rapidity.}
\label{fig:6}
\end{figure}

A boost-invariant way to display these results is on the $r_{\perp} - \eta$ plane, as shown in Fig. \ref{fig:6}, but there the volumes involved are not apparent. 

It should be emphasized that the baryon densities calculated here are more robust than the energy densities.  The reason can be seen by comparing Figs. \ref{fig:2} and \ref{fig:3}.  The rapidity loss, and therefore compression, are determined mostly within the first few tenths of a fm/c when the glasma dominates the dynamics.  The excitation energy continues its slow growth as time goes on.  If the transition from glasma to quark-gluon plasma happens earlier than 0.6 fm/c it would reduce the excitation energy but hardly affect the compression.  Exactly how the transition occurs is a topic of much current interest and activity.  Possibilities include: instabilities due to initial state fluctuations \cite{RV,FujiiItakura,KM,Gelis:2013rba}, a universal attractor solution which governs the late time evolution in the classical regime \cite{Berges}, and rapid conversion of classical fields to partons with subsequent evolution of the system described by a Boltzmann equation \cite{KZ2015}.  This should be kept in mind in the following discussions.

The above results do not assume that the fireballs thermalize.  Now, out of curiosity, let us assume that the matter in the fireballs does equilibrate on the time scale of 0.6 fm/c as argued in Ref. {\cite{AKM1980}.  What does that imply for the temperatures and chemical potentials attainable?  That requires an equation of state.  For simplicity consider a massless gas of noninteracting up, down, and strange quarks and gluons.  (A recent QCD perturbative calculation of the equation of state at high chemical potential \cite{KV2016} would give similar results.)  The net strangeness in the fireball is zero which means that the chemical potential of the strange quark is zero.  Let the up and down quark chemical potentials be equal to each other and to 1/3 of the baryon chemical potential $\mu_{\rm B}$.  Then the charge to baryon ratio is 0.5 versus 0.4 for a gold nucleus.  The equation of state is then given by the pressure as a function of temperature and baryon chemical potential
\begin{equation}
P(T,\mu_{\rm B}) = \frac{19 \pi^2}{36} T^4 + \frac{1}{9} T^2 \mu_{\rm B}^2 + \frac{1}{162 \pi^2} \mu_{\rm B}^4 \,.
\end{equation}
Taking the energy density 20 GeV/fm$^3$ and the baryon density 3 baryons/fm$^3$ (20 times normal nuclear matter density) results in a temperature $T = 299$ MeV and $\mu_{\rm B} = 1061$ MeV (approximating $\hbar c = 200$ MeV$\cdot$fm).  Thus the up and down quark chemical potentials are greater than the temperature, very unlike in the central rapidity region at the top RHIC energy or at the LHC.  The entropy per baryon of 26.2 is still quite large.  However, the energy density decreases with $r_{\perp}$ faster than the baryon density, as can be seen in Fig. \ref{fig:4}.  Taking the energy density 5.5 GeV/fm$^3$ and the baryon density 1.5 baryons/fm$^3$ results in $T = 205$ MeV and $\mu_{\rm B} = 1007$ MeV with an entropy per baryon of 18.9.  

\begin{figure}[b]
\includegraphics[width=0.999\columnwidth]{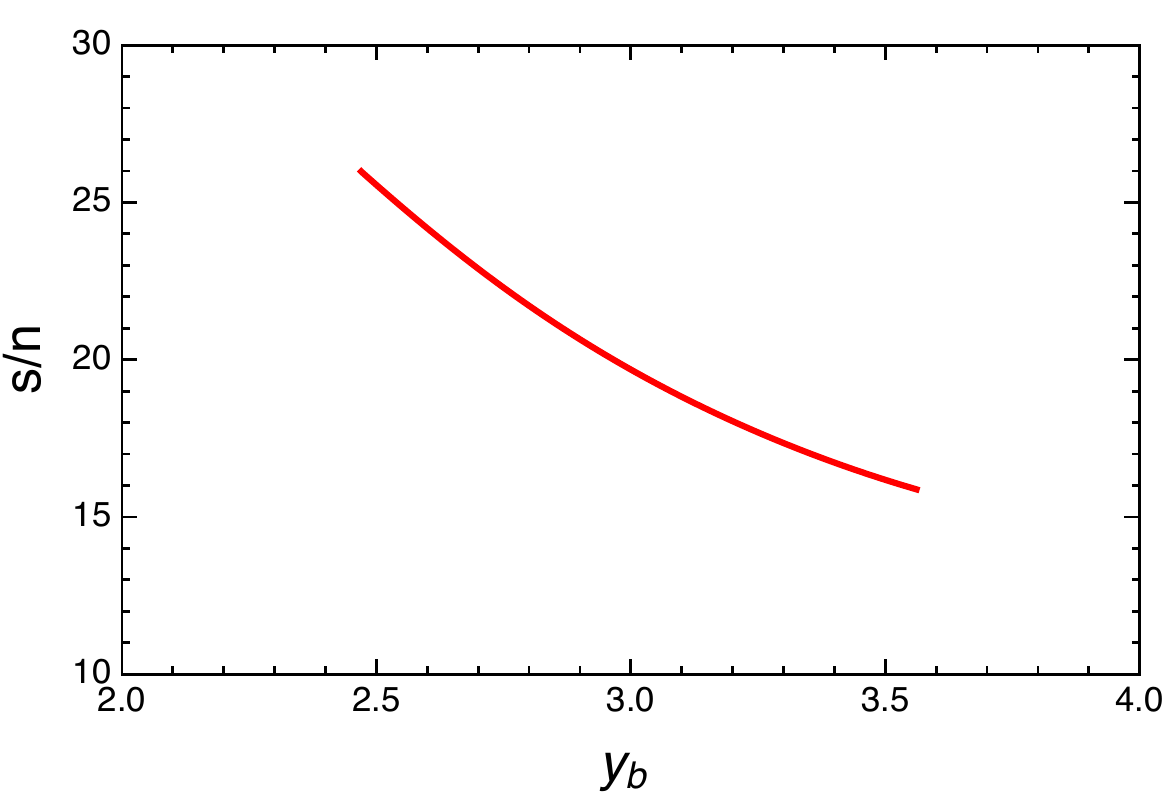}
\caption{\label{F6}(color online) Entropy per baryon rapidity distribution at $\tau = 0.6$ fm/c. A rapidity scan might help locate a critical point.} 
\label{fig:7}
\end{figure}
It would be expected that the hydrodynamic expansion of the fireball would be approximately adiabatic, just as in the central rapidity region.  If that is the case, then the values of the entropy per baryon estimated above would be in just the right range for the trajectories of the fluid elements on the $T-\mu_{\rm B}$ plane to pass near or even through a possible critical point in the QCD phase diagram \cite{ABMN2008}.   Figure \ref{fig:7} shows the entropy per baryon as a function of rapidity, and begs the question of whether a rapidity scan could help locate a possible critical point.

In a follow-up paper we will study the systematics of high baryon densities achievable in high energy heavy ion collisions, such as the dependence on impact parameter, beam energy, nuclear size, projectiles and targets of different mass, and so on.  Beyond that, one must consider that we are only proposing initial conditions for subsequent hydrodynamic evolution of the hot and dense fireball.  This cannot be studied on its own but must incorporate the production of quark-gluon plasma in the region between the receding fireballs.  Generally this will broaden the baryon rapidity distribution due to collective flow and thermal smearing in the final state.

As mentioned earlier, detectors at RHIC and LHC that are or were being used for heavy ion collisions focus on central rapidities, generally within a few units of $y=0$.  Apart from BRAHMS, which was still limited to about three units of rapidity, this precludes them from studying the range of rapidities of the hadrons emerging from the fireballs.  Even if the LHC was to operate in a fixed target mode at a beam energy of 2.76 GeV per nucleon for lead nuclei, this would only provide a $\sqrt{s_{\rm NN}}$ of 72 GeV, which is already within the RHIC energy range and near the lower limit of applicability of the McLerran-Venugopalan model.  

In conclusion, we have employed the McLerran-Venugopalan model to calculate the energy/rapidity loss of baryons in high energy heavy ion collisions.  Very similar results should be obtained in different pictures of heavy ion collisions, even though the language is rather different \cite{Chesler1,Chesler2,PENT2014}.  We found that the baryon densities in the fireballs outside the central rapidity region attain values an order of magnitude greater than normal nuclear matter.  These findings suggest that further theoretical and experimental studies be performed to probe the equation of state at the highest baryon densities achievable in a laboratory setting.

\section*{Acknowledgement}
We are grateful to L. McLerran and B. M\"uller for comments on the manuscript.  This work was supported by the U.S. Department of Energy Grant  DE-FG02-87ER40328.

\end{document}